\documentclass[aps,eqsecnum,nofootinbib,showpacs]{revtex4}
\usepackage{amsmath,amsfonts,bm,amssymb}

\newcommand{\be}{\begin{equation}}
\newcommand{\ee}{\end{equation}}
\newcommand{\bea}{\begin{eqnarray}}
\newcommand{\eea}{\end{eqnarray}}

\newcommand{\eq}[1]{Eq.~(\ref{eq:#1})}
\newcommand{\sect}[1]{Sec.~\ref{sec:#1}}

\newcommand{\tpi}{\tau_{\pi}}
\newcommand{\tpishear}{\tpi^{\rm (shear)}}
\newcommand{\tpisound}{\tpi^{\rm (sound)}}
\newcommand{\tPi}{\tau_{\Pi}}

\newcommand{\Ks}{K_*}
\newcommand{\deltakt}{\delta_{\rm KT}}
\newcommand{\deltadp}{\delta_{Dp}}
\newcommand{\ft}[2]{{\textstyle{\frac{#1}{#2}}}}

\begin{document}


%
\title{
Causal hydrodynamics and the membrane paradigm}
\author{Makoto Natsuume}
\email{makoto.natsuume@kek.jp}
\affiliation{Theory Division, Institute of Particle and Nuclear Studies, \\
KEK, High Energy Accelerator Research Organization, Tsukuba, Ibaraki, 305-0801, Japan}
\date{\today}
\begin{abstract}
We obtain the relaxation time for the shear viscous stress for various geometries using the ``membrane paradigm" formula proposed recently.  We consider the generic Schwarzschild-AdS black holes (SAdS), the generic D$p$-brane, the Klebanov-Tseytlin (KT) geometry, and the ${\cal N}=2^*$ theory. The formula is the ``shear mode" result and is not fully trustable, but it may be helpful to learn some generic behaviors about the relaxation time. For example, a simple formula  summarizes all known results for SAdS, and a single expression summarizes the results for the D$p$-brane and the KT geometry.
\end{abstract}

\maketitle

\section{Introduction}

The AdS/CFT duality is a powerful tool to study hydrodynamics of gauge theory plasmas.%
\footnote{
See Ref.~\cite{Natsuume:2007qq} for a review.}
However, standard hydrodynamics (first-order formalism) has severe problems such as acausality. 
One can restore causality by introducing a new set of transport coefficients, and the resulting theory is known as ``causal hydrodynamics"  or  ``second-order formalism." In modern language, it is just an effective theory expansion in higher orders. At present, there is no unique formalism for causal hydrodynamics, but probably the most used formalism is the ``Israel-Stewart theory" \cite{Imuller,Israel:1976tn,Israel:1979wp}.  Causal hydrodynamics has been widely discussed in the context of heavy-ion collisions. 

Recently, a number of papers appeared which study the causal hydrodynamics of gauge theory plasmas using the AdS/CFT duality \cite{Natsuume:2007ty}-\cite{Bhattacharyya:2007jc}.%
\footnote{See also Refs.~\cite{Natsuume:2008iy}-\cite{Kinoshita:2008dq} and Ref.~\cite{causal_review} for a review.}
One main focus of these works is to determine the new transport coefficients, in particular, the relaxation time $\tpi$ for the shear viscous stress. 

There are several methods to obtain transport coefficients in the AdS/CFT duality, but the ``membrane paradigm" method is the simplest and the most powerful one \cite{Kovtun:2003wp}. This method was helpful to establish the universality of $\eta/s$ for a broad range of gauge theories, where $\eta$ and $s$ are the shear viscosity and the entropy density, respectively. Reference~\cite{Kapusta:2008ng} has proposed the membrane paradigm formula for the relaxation time {\it from the shear mode}.%
\footnote{
The gravitational perturbation or the energy-momentum tensor can be decomposed as the tensor mode, the vector mode (``shear mode"), and the scalar mode (``sound mode"). 
}

Using the membrane paradigm formula, we estimate the relaxation time for various geometries:
\begin{enumerate}
\item Conformal theories: the generic $(p+2)$-dimensional Schwarzschild-AdS black hole (SAdS$_{p+2}$) 
\item Nonconformal theories:
\begin{enumerate}
\item The generic D$p$-brane \label{item1}
\item The Klebanov-Tseytlin geometry \label{item2}
\item The ${\cal N}=2^*$ theory \label{item3}
\end{enumerate}
\end{enumerate}
One should be careful to interpret these results. As argued in Refs.~\cite{Natsuume:2007ty}-\cite{Bhattacharyya:2007jc}, the relaxation time from the shear mode is unreliable. The relaxation time from the sound mode is reliable, but the sound mode computations are often harder due to the lack of symmetry. Since the membrane paradigm method is a shear mode method, the result is not trustable. 

It is not our aim here to determine the correct coefficients however. Rather, we use the formula to learn some generic features about the relaxation time. (This may be possible since the functional forms are often similar in both modes.) For example, a simple formula by a harmonic number summarizes all known results for SAdS backgrounds. Also, it is known that the relaxation time is not the same in different spacetime dimensions, but our analysis indicates that the relaxation time is not the same even in a given spacetime dimension. Additionally, the relaxation time for geometry~\ref{item1} and \ref{item2} is written in the same form by the speed of sound, but this is not the case for geometry~\ref{item3}.
Currently, little is known about the relaxation time $\tpi$ for nonconformal theories since the sound mode is not enough to determine $\tpi$: another relaxation time $\tPi$ appears in the sound mode, so one cannot determine $\tpi$ and $\tPi$ separately. 
We hope that some results presented here will be justified from the sound mode or will give a clue for solving the sound mode in the future.

\section{Membrane paradigm formula}

The membrane paradigm formula for the relaxation time {\it from the shear mode} is given by \cite{Kapusta:2008ng}
\be
\tpishear = \frac{\sqrt{-g(r_0)}}{\sqrt{-g_{00}(r_0) g_{rr}(r_0)}}
\int_{r_0}^\infty dr \frac{g_{rr}(r)}{\sqrt{-g(r)}}
\left[ 1- \left( \frac{D(r)}{D(r_0)} \right)^2 \right]~,
\label{eq:membrane_formula}
\ee
where
\be
D(r) := \frac{\sqrt{-g(r)}}{\sqrt{-g_{00}(r)g_{rr}(r)}}
\int_r^\infty dr'
\frac{-g_{00}(r')g_{rr}(r')}
{\sqrt{-g(r')}g_{xx}(r')}~.
\label{eq:nested_integral}
\ee
Here, $r=r_0$ represents the location of the black hole horizon.
The nested integral $D(r_0)$ is just the membrane paradigm formula derived in Ref.~\cite{Kovtun:2003wp} for a diffusion constant $D_\eta := \eta/(T+s)$.
One can use either the 10-dimensional metric or the compactified metric, but one should use the Einstein frame. 
Note that the formula is invariant under the change of the radial coordinate $r$. To derive \eq{membrane_formula}, they used SAdS$_5$-like radial coordinate $r$. But it is sometimes more convenient to use a coordinate other than $r$ (See, {\it e.g.}, \sect{kt} and \sect{PW}). The formula is not affected by the change of the radial coordinate since $g_{rr}$ component appears only in the forms of $g_{rr}\, dr/\sqrt{-g}$ and $\sqrt{-g/g_{rr}}$. So, one can choose a radial coordinate at will. 

{\it One should not take the formula too literally.} As discussed in Refs.~\cite{Natsuume:2007ty}-\cite{Bhattacharyya:2007jc}, the relaxation time can be determined both from the shear mode and from the sound mode. However, the shear mode result is unreliable due to the contamination from the ``third-order hydrodynamics." On the other hand, the sound mode result is free from this problem. The coefficient $\tpishear$ should be understood as a quantity to summarize gravity results (including the possible contamination from the third-order hydrodynamics).

\section{Conformal theories: ${\rm SAdS}_{p+2}$}

The ${\rm SAdS}_{p+2}$ background is dual to a $(p+1)$-dimensional conformal theory. The metric is given by
\be
ds_{p+2}^2 = 
f (-hdt^2+d\vec{x}_p^2) +  \frac{dr^2}{fh}~,
\label{eq:sads_metric}
\ee
where 
\bea
f &=& \left( \frac{r}{R} \right)^2~, \\
h &=& 1-\left( \frac{r_0}{r} \right)^{p+1}~,
\eea 
where $R$ is the AdS radius.
The temperature is given by
\be
4\pi T=(p+1)\frac{r_0}{R^2}~.
\ee

From \eq{membrane_formula}, we obtain
\be
(4\pi T)\, \tpishear = H_{\frac{2}{p+1}}~,
\label{eq:shear_sads}
\ee
where $H_n$ is a harmonic number. For our purpose, it is useful to use an integral representation of the harmonic number:
\be
H_n = \int_0^1 dx\, \frac{1-x^n}{1-x}~.
\ee
A harmonic number can also be written as $H_n = \gamma+\psi(n+1)$, where $\gamma$ and $\psi(n)$ are the Euler constant and a digamma function, respectively. Equation~(\ref{eq:shear_sads}) reproduces all known results for SAdS$_{p+2}$ ($p=2,3,5$) \cite{Natsuume:2007ty,Baier:2007ix,Bhattacharyya:2007jc}. For the SAdS$_6$,
\be
(4\pi T)\, \tpishear = \frac{5}{2}-\frac{\pi}{2}\sqrt{1-\frac{2}{\sqrt{5}}}+\frac{\sqrt{5}}{2}\coth^{-1}\sqrt{5}-\frac{5\ln5}{4}~.
\ee
The relaxation time (\ref{eq:shear_sads}) is monotonically decreasing with $p$, but this feature is misleading. The results from the shear mode is unreliable; in fact, the results from the sound mode suggests that the relaxation time is monotonically increasing with $p$ [See \eq{sound_sads}]. 

Combining the result (\ref{eq:shear_sads}) with known results in the sound mode, one is tempted to conjecture a formula in the sound mode:
\be
(4\pi T)\,\tpisound  = H_{\frac{2}{p+1}}+\frac{p+1}{2}~,
\label{eq:sound_sads}
\ee
which reproduces all known results for SAdS$_{p+2}$ ($p=2,3,4,5$) \cite{Natsuume:2007ty,Baier:2007ix,Bhattacharyya:2007jc,Haack:2008cp}, but it is not currently clear if \eq{sound_sads} is valid for a generic $p$. Let us suppose that it is indeed valid for a generic $p$ and consider the $p\rightarrow\infty$ asymptotic behavior. In causal hydrodynamics, the signal propagation speed $v_{\rm \,front} $ may be defined as 
\be
v_{\rm \,front}^2 := \frac{D_\eta}{ \tpisound }, \quad
\mbox{where}\quad
D_\eta :=\frac{\eta}{\epsilon+p}~.
\ee 
Then, the signal propagation speed asymptotically behaves as $v_{\rm \,front} \sim \sqrt{2/p}$ for a large $p$, whereas the speed of sound behaves as $v_s =\sqrt{1/p}$.

\section{Nonconformal theories}

For nonconformal theories, the correct value of the relaxation time $\tpi$ from the sound mode is so far unknown; at least one other parameter, another relaxation time $\tPi$ appears in the sound mode, so one cannot determine $\tpi$ and $\tPi$ separately. A combination of $\tpi$ and $\tPi$ has been determined for the D4-brane in Ref.~\cite{Natsuume:2007ty}.

\subsection{D$p$-brane}

In the ``near-horizon" limit, the D$p$-brane is dual to the $(p+1)$-dimensional SYM with 16 supercharges. 
The 10-dimensional Einstein metric (for $p<7$) is given by 
\be
ds_{\rm E}^2 =
 Z^{-\frac{7-p}{8}} (-hdt^2+d\vec{x}_p^2) 
 + Z^{\frac{p+1}{8}}  (h^{-1}dr^2+r^2 d\Omega_{8-p}^2) ~,
\label{eq:dp_metric} 
\ee
where 
\bea
Z &=& \left( \frac{r}{R} \right)^{-(7-p)}~, \\
h &=& 1-\left( \frac{r_0}{r} \right)^{7-p}~.
\eea 
The temperature is given by
\be
4\pi T = (7-p)\frac{r_0^{\frac{5-p}{2}}}{R^{\frac{7-p}{2}}}~.
\ee

From \eq{membrane_formula}, we obtain
\be
(4\pi T)\, \tpishear 
= H_{\frac{5-p}{7-p}}~
\label{eq:shear_Dp}
\ee
for $p<5$. Equation~(\ref{eq:shear_Dp}) reproduces all known results for the D$p$-brane ($p=3,4$) \cite{Natsuume:2007ty,Baier:2007ix,Bhattacharyya:2007jc}. 
For the D2-brane,
\be
(4\pi T)\,\tpishear = \frac{5}{3}+\frac{\pi}{2}\sqrt{1-\frac{2}{\sqrt{5}}}+\frac{\sqrt{5}}{2}\coth^{-1}\sqrt{5}-\frac{5\ln5}{4}~.
\ee
For the D5-brane, \eq{membrane_formula} vanishes, but some of the intermediate expressions to derive \eq{membrane_formula} actually diverge in this case.
For the D6-brane, \eq{membrane_formula} diverges. This implies that the second-order corrections are large so that the second-order theory is not useful. These peculiar behaviors may be related to the well-known instability of the D$p$-brane for $p \ge 5$. The specific heat $C$ and the speed of sound $v_s$ are given by
\be
C = \frac{\partial \epsilon}{\partial T} = \frac{9-p}{5-p}\, s~, \qquad
v_s^2 = \frac{5-p}{9-p}~,
\ee
where $s$ is the entropy density. For $p=5$, the specific heat diverges and the speed of sound vanishes; for $p=6$, the specific heat becomes negative and the speed of sound becomes imaginary.

If one regards
\be
\deltadp:= \frac{p-3}{4}
\ee
as a deformation parameter from the conformal $p=3$ theory, 
\bea
\tpishear T 
  &\sim& H_{\frac{1-\deltadp}{2}} \\
  &=& \frac{1-\ln2}{2\pi} - \frac{\pi^2-8}{16\pi}\,\deltadp + O(\delta^2)~.
\eea
If one can interpret this case as a nonconformal theory in 4 dimensions, this suggests that the relaxation time deviates from the ${\cal N}=4$ result as one deviates from the conformal theory.%
\footnote{The D$(3+\epsilon)$-brane has been discussed in Refs.~\cite{Caceres:2006ta,Natsuume:2007vc}. See also Ref.~\cite{Alday:2007hr} for a discussion somewhat in a different context.}

\subsection{Klebanov-Tseytlin geometry}\label{sec:kt}

The Klebanov-Tseytlin geometry is dual to the ${\cal N}=1$ cascading $SU(\Ks) \times SU(\Ks+P)$ gauge theory. For temperatures high above the
deconfining transition, the solution was constructed in Refs.~\cite{Gubser:2001ri,Buchel:2001gw,Buchel:2000ch}. In this regime, the theory is parametrized by the deformation parameter $\deltakt$:
\be
\deltakt := \frac{P^2}{\Ks} \ll 1~.
\ee
The 10-dimensional Einstein metric involves 3 functions $\xi(z)$, $\eta(z)$, and $\omega(z)$ of a radial coordinate $z$. To leading order in $\deltakt$, the solution is given by
\begin{eqnarray}
ds^2_{\rm E} &=& 
{\sqrt{8a/\Ks}\over\sqrt{z}}e^{2P^2\eta}
\left(-(1-z)dt^2+d\vec{x}_3^2\right)
+ {\sqrt{\Ks}\over32}e^{-2P^2(\eta-5\xi)}{dz^2\over z^2(1-z)}
  \nonumber \\
{}&{}&\qquad\qquad
+{\sqrt{\Ks}\over 2}e^{-2P^2(\eta-\xi)}
\left[e^{-8P^2\omega}e^2_{\psi}
+e^{2P^2\omega}(e^2_{\theta_1}+e^2_{\phi_1}+e^2_{\theta_2}+e^2_{\phi_2})\right]~,
  \\
\xi&=&{2z+[-2z+(z-2)\ln(1-z)]\ln z+(z-2){\rm Li}_2(z)\over 40 \Ks z}~, \\
\eta&=&{ z-2 \over 16 \Ks z} [\ln z \ln (1-z) +{\rm Li}_2 (z)]~, 
%
\end{eqnarray}
where ${\rm Li}_2 (z)$ is a polylogarithm. The explicit form of $\omega(z)$ is not necessary for our purpose; the volume of the compact space does not depend on it, and as a consequence, it does not appear in \eq{membrane_formula}. Also, $z \in [0,1]$, with $z=1$ corresponding to the location of the horizon.
The temperature is determined by the standard Euclidean continuation:
\be
T= \frac{2\, (2a)^{1/4}}{\sqrt{\Ks} \pi} e^{-\frac{P^2}{4\Ks}}~.
\ee

From \eq{membrane_formula}, we obtain
\be
(4\pi T)\, \tpishear 
= H_{\frac{1-\deltakt}{2}}
\ee
or
\be
\tpishear T 
  = \frac{1-\ln2}{2\pi} - \frac{\pi^2-8}{16\pi}\,\deltakt + O(\delta^2)~,
\ee
which takes the same form as the D$p$ result. (See \sect{discussion}.)

\subsection{${\cal N}=2^*$ theory}\label{sec:PW}

The ${\cal N}=2^*$ $SU(N)$
gauge theory is parametrized by the deformation parameters
\be
\alpha_1\propto \left(\frac{m_b}{T}\right)^2 \ll 1~, \qquad
\alpha_2\propto \frac{m_f}{T} \ll 1~, 
\ee
where $m_b$ and $m_f$ are masses of the bosonic and fermionic 
components of the ${\cal N}=2$ hypermultiplet. The solution was constructed in Ref.~\cite{Buchel:2003ah}.

The 5-dimensional Einstein metric is of the form
\be
ds^2_{\rm E}  = e^{2A} \left( - e^{2 B} dt^2 + d\vec{x}_3^2 \right) + dr^2~.
\ee
Following Ref.~\cite{Buchel:2003ah}, introduce a new radial coordinate, $y=e^B$, $y\in [0,1]$, with $y=0$ corresponding to the location of the horizon.
Then, the metric becomes
\be
ds^2_{\rm E} =
 e^{2A} \left( - y^2 dt^2 + d\vec{x}_3^2\right)
  + dy^2 \left(\frac{\partial r}{\partial y}\right)^2~. 
\ee
To the leading order in $\alpha_1$ and $\alpha_2$, the solution 
is 
\bea
A(y) &=& \hat\alpha - \frac{1}{4} \ln (1-y^2) 
                 + \alpha_1^2 A_1(y) + \alpha_2^2 A_2(y)~,\\
\rho(y) &=& 1 + \alpha_1 \rho_1(y)~,\\
\chi(y) &=& \alpha_2 \chi_2(y)~,
\eea
where $\rho$ and $\chi$ are two scalars whose solutions are
\bea
\rho_1 &=&(1-y^2)^{1/2}\ _2F_1(\ft{1}{2}, \ft{1}{2}, 1; y^2)~, \\
\chi_2 &=& (1-y^2)^{3/4}\ _2F_1(\ft{3}{4}, \ft{3}{4}, 1;y^2)~,
\eea
and
\bea
A_1 &=& - 4 \int_0^y\, \frac{z\,dz}{(1-z^2)^2}\
\biggl( \gamma_1
+\int_0^z dx\,\left( \frac{\partial\rho_1}{\partial x} \right)^2\,
\frac{(1-x^2)^2 }{x}\biggr)~,
\\
A_2 &=& -\frac{4}{3} \int_0^y\, \frac{z\,dz}{(1-z^2)^2}\
\biggl( \gamma_2
+\int_0^z dx\,\left( \frac{\partial\chi_2}{\partial x} \right)^2\,
\frac{(1-x^2)^2 }{x}\biggr)~.
\eea
The constants $\gamma_i$ are given by
\be
\gamma_1 = \frac{8-\pi^2}{2 \pi^2}~, \qquad
\gamma_2 = \frac{8-3\pi}{8 \pi}~.
\ee
The parameters $\alpha_i$ are related to the parameters $m_b$, $m_f$, $T$, and $v_s$ of the dual gauge theory via \cite{Buchel:2003ah,Benincasa:2005iv}
\bea
\alpha_1 &=& -\frac{1}{24\pi}\ \left(\frac{m_b}{T}\right)^2~,\\
\alpha_2 &=& \frac{ \left[ \Gamma\left(\ft 34\right) \right]^2 }{2\pi^{3/2}}\ 
\frac{m_f}{T}~,\\
2\pi T &=& e^{\hat\alpha}
\left(1+\frac{16}{\pi^2} \alpha_1^2+\frac{4}{3\pi} \alpha_2^2 \right)~,\\
3 v_s^2 &=& 1 - \frac{64}{\pi^2} \alpha_1^2 - \frac{8}{3\pi} \alpha_2^2~.
\label{eq:PW_sound_speed}
\eea

The integral (\ref{eq:nested_integral}) gives
\be
D(y) = e^{3 A(y)} \int_y^1 dy'\,
\frac{ y' e^{-4 A(y')} }{ |\frac{\partial y'}{\partial r}| }~. 
\ee
Following an argument of Appendix~C in Ref.~\cite{Kovtun:2003wp}, expand the Jacobian to the leading order in $\alpha_1$ and $\alpha_2$, and one can show
\be
\frac{ e^{-4 A(y')} }{ |\frac{\partial y'}{\partial r}| } = e^{-3\hat\alpha} \frac{1}{2\pi T}+\cdots~,
\label{eq:jacobian}
\ee
where the dots stand for terms of higher orders. Then,
\be
D(y) = \frac{1}{4\pi T} e^{3 (A-\hat\alpha)} (1-y^2)~. 
\ee
Thus, the relaxation time is given by
\be
\tpishear = e^{3 \hat\alpha} \int_0^1 dy\,
\frac{1}{y}
\frac{ e^{-4 A} }{ |\frac{\partial y}{\partial r}| }
[ 1-e^{6 (A-\hat\alpha)}(1-y^2)^2 ]~.
\ee
Using \eq{jacobian} again, we get
\bea
(2\pi T) \tpishear 
&\sim& \int_0^1 dy\, \frac{1}{y}[1-(1-y^2)^{1/2}]
- 6 (\alpha_1^2 I_1 + \alpha_2^2 I_2) \\
&=& (1 - \ln2) - 6 (\alpha_1^2 I_1 + \alpha_2^2 I_2)~,
\eea
where
\be
I_i := \int_0^1 dy\, \frac{(1-y^2)^{1/2}}{y} A_i~.
\ee
The integral $I_1$ is written as
\be
I_1= -2\gamma_1 
- 2 \int_0^1 dy\, \left[ \frac{(1-y^2)^{3/2}}{y} + y(1-y^2) \ln\left( \frac{y}{1+\sqrt{1-y^2}} \right) \right](\rho_1')^2
\label{eq:PW_final_form}
\ee
by integrations by parts. 
The integral $I_2$ can be written similarly. 
We are not able to obtain analytical expressions for integrals $I_i$, but it is easy to estimate them numerically:
\be
\tpishear T \sim \frac{1-\ln2}{2\pi} - 0.0733\, \alpha_1^2 - 0.0151\, \alpha_2^2~ + O(\alpha^4)~.
\label{eq:PWresult}
\ee

\section{Discussion}\label{sec:discussion}

The relaxation time is not the same in different spacetime dimensions from Ref.~\cite{Natsuume:2007ty}, but there is a possibility that the relaxation time is the same in each spacetime dimension. Our analysis indicates that this is not the case: the relaxation time depends on the deformation parameter. 

On the other hand, we found that the results for the D$p$-brane and the KT geometry take the same form. For the D$p$-brane and the KT geometry, the speed of sound $v_s$ is given by a single formula if one uses our definition for deformation parameters \cite{Caceres:2006ta,Buchel:2005cv}:
\be
v_s^2 = \frac{1}{3}\left(1-\frac{4}{3} \delta\right)
+O\left(\delta^2\right)~.
\label{eq:vsbydelta}
\ee
Then, one can rewrite the nonconformal corrections to the relaxation time by the speed of sound:
\be
\delta\tpishear T = - \frac{3(\pi^2-8)}{64\pi}(1-3v_s^2)
+O\left((1-3v_s^2)\right)~.
\label{eq:tau_by_vs}
\ee
Let us make a simple estimate of the correction for QCD. If we use 1 fm for the inverse temperature, the conformal result is $\tpisound \sim 0.2\, {\rm fm}$. According to the lattice results cited in Ref.~\cite{Karsch:2006sm}, all groups roughly predict $1/3-v_s^2 \sim 0.05$ around $2T_c$. Bearing in mind that our results are valid to large-$N_c$ theories and not to QCD, the nonconformal correction decreases the relaxation time about 2\%.

However, the ${\cal N}=2^*$ theory does not take the same form as the above geometries. To compare with the above geometries, define new parameters $\delta_i$ as
\be
\delta_1 := \frac{48}{\pi^2} \alpha_1^2~, \qquad
\delta_2 := \frac{2}{\pi} \alpha_2^2~.
\ee
Then, the speed of sound (\ref{eq:PW_sound_speed}) takes the same form as \eq{vsbydelta}, {\it i.e.},
\be
v_s^2 = \frac{1}{3}\left(1-\frac{4}{3} (\delta_1+\delta_2) \right)
+O\left(\delta^2\right)~.
\ee
However, the correction to the relaxation time (\ref{eq:PWresult}) is
\be
\delta\tpishear T \sim - 0.0151\, \delta_1 - 0.0237\, \delta_2  + O(\delta^2)~,
\ee
whereas the relation for the D$p$-brane and the KT geometry is (numerically) 
\be
\delta\tpishear T \sim - 0.0372\, \delta_{Dp, {\rm KT}} + O(\delta^2)~.
\ee
Thus, \eq{tau_by_vs}  does not seem a universal expression. 

It is a curious fact that the D$p$-brane and the KT geometry often satisfy the same expressions, but it is not well understood at the moment. For example, these geometries satisfy the same expression for the bulk viscosity but the ${\cal N}=2^*$ theory does not \cite{Benincasa:2005iv,Buchel:2005cv,Mas:2007ng}. Reference~\cite{Caceres:2006ta} also observed that the D$p$-brane and the KT geometry satisfy the same expression in the context of screening length.

\begin{acknowledgments}
I would like to thank Takashi Okamura and Todd Springer for useful discussions. This work was supported in part by the Grant-in-Aid for Scientific Research (20540285) from the Ministry of Education, Culture, Sports, Science and Technology, Japan.
\end{acknowledgments}

\end{document}